\documentclass{aa}
\usepackage{graphicx}
\usepackage{txfonts}
\begin{document}
   \title{Empirical strong-line
oxygen abundance calibrations from galaxies with electron temperature
measurements}

   \titlerunning{Oxygen abundance calibrations}

   \author{S. Y. Yin\inst{1,3}, Y. C. Liang\inst{1,2}, F. Hammer\inst{2},
          J. Brinchmann\inst{4},
          B. Zhang\inst{1,3}, L. C. Deng\inst{1}, and H. Flores\inst{2}
           }
    \authorrunning{Yin et al.}

   \offprints{S. Y. Yin: syyin@bao.ac.cn,  Y. C. Liang: ycliang@bao.ac.cn}

   \institute{
  $^1$National Astronomical Observatories, Chinese Academy of Sciences,
     20A Datun Road, Chaoyang District, Beijing 100012, China\\
  $^2$GEPI, Observatoire de Paris-Meudon, 92195 Meudon, France \\
  $^3$Department of Physicals,Hebei Normal University,Shijiazhuang 050016, China \\
  $^4$ CAUP, Rua das Estrelas S/N, 4150-752 Porto, Portugal
 }

 \date{Received; accepted}


  \abstract
   { }
   {We determine the gas-phase oxygen abundance for a sample
       of 695 galaxies and H~\textsc{ii} regions with reliable detections of
       [O~\textsc{iii}]4363, using the temperature-sensitive ($T_e$) method,
      which is the most reliable and direct way of measuring metallicity.
    Our aims are to estimate the validity of empirical methods such
     as $R_{23}$, $R_{23}-P$, log([N {\sc ii}]/H$\alpha$) (N2),
     log[([O {\sc iii}]/H$\beta$)/([N {\sc ii}]/H$\alpha$)] (O3N2),
     and log([S {\sc ii}]/H$\alpha$) (S2), and re-derive (or add) the
     calibrations of $R_{23}$, N2, O3N2 and S2 indices for oxygen abundances on
     the basis of this large sample of galaxies with $T_e$-based abundances.  }
   {We select 531 star-forming galaxies from the Fourth Data
     Release of the Sloan Digital Sky Survey database (SDSS-DR4) with
     strong emission lines, including [O~{\sc iii}]4363
     detected at a signal-to-noise larger than 5$\sigma$, as well as
     164 galaxies and H~\textsc{ii} regions from literature with $T_e$
       measurements. O/H abundances have been derived from a
     two-zone model for the temperature structure, assuming a
     relationship between high ionization and low ionization species.
   }
   {  We compare our (O/H)$_{T_e}$ measurements of the SDSS sample with
 the abundances obtained by the MPA/JHU group using multiple strong
 emission lines and Bayesian techniques (Tremonti et al. 2004).  For
 roughly half of the sample the Bayesian abundances are overestimated
 $\sim$0.34\,dex, possibly due to the treatment of nitrogen enrichment
 in the models they used.
        $R_{23}$
     and $R_{23}-P$ methods systematically overestimate the O/H abundance by a
     factor of $\sim$0.20\,dex and $\sim$0.06\,dex, respectively.  The
     N2 index, rather than the O3N2 index,
     provides more consistent O/H abundances with the
     $T_e$-method, but with some scatter. The relations of N2,
     O3N2, S2 with log(O/H) are consistent with the photoionization
     model calculations of Kewley \& Doptita (2002),
     but $R_{23}$ does not match well.
     We derive analytical calibrations for O/H from $R_{23}$,
     N2, O3N2 and S2 indices on the basis of this large sample,
     as well including the excitation parameter $P$ as an
     additional parameter in the N2 calibration.
     These empirical calibrations are free from the systematic problems
     inherent in abundance calibrations based on photoionizatoin models.
      }
   { We conclude that,
   the N2, O3N2 and S2 indices are useful indicators
   to calibrate metallicities of galaxies with 12 + log(O/H) $< $ 8.5,
   and the $R_{23}$ index works well for the metal-poor galaxies
   with 12+log(O/H)$<$7.9. For the intermediate metallicity
     range (7.9 $<$ 12 + log(O/H) $< $ 8.4), the $R_{23}$ and $R_{23}-P$ methods
     are unreliable to characterize the O/H abundances.
 }

   \keywords{galaxies: abundances -
          galaxies: evolution -
          galaxies: ISM -
          galaxies: spiral -
          galaxies: starburst -
          galaxies: stellar content}
   \maketitle
%

%
\section{Introduction}
The chemical properties of stars and gas within a galaxy
provides both a fossil record of its star formation history
  and information on its present evolutionary status. It is therefore
  desirable to extract as much, and as accurate, information as
  possible from observations of galaxies. In particular it is
  important that different methods for extracting information provides
  this without systematic offsets, or at the very least that these
  systematic offsets are well understood. Accurate abundance
measurements for the ionized gas in galaxies require the
determination of the electron temperature ($T_e$) in this gas
which is usually obtained from the ratio of auroral to nebular
line intensities, such as [O~{\sc
iii}]$\lambda\lambda4959,5007$/[O~{\sc iii}]$\lambda$4363. This is
generally known as the ``direct $T_e$-method" because the electron
temperature is
  directly inferred from observed line ratios. It is well known that
  this procedure is difficult to carry out for metal-rich galaxies
  since, as the metallicity increases, the electron temperature
decreases (as the cooling is via metal lines) and the auroral lines
eventually become too faint to measure. Instead, the most common
method used for estimating oxygen abundance of metal-rich galaxies
(12+log(O/H)$\geq$8.5) uses the $R_{23}$ (=([O~{\sc
  ii}]$\lambda$3727+[O~{\sc iii}]$\lambda\lambda$4959,5007)/H$\beta$)
parameter, which is the ratio of the flux in the strong
  optical oxygen lines to that of H$\beta$ (Pagel et
al. 1979; Tremonti et al. 2004 and the references therein).  The
$R_{23}$ indicator can also be used for metal-poor galaxies
(12+log(O/H)$<$8.5) (Skillman et al. 1989; Kobulnicky et al. 1999;
McGaugh 1991; Pilyugin 2000; Edmunds \& Pagel 1984).

Several researchers have however found that $R_{23}$-derived
  abundances are inconsistent with the $T_e$-derived ones, showing a
  systematic offset. For example, Kennicutt et al. (2003) found that
$R_{23}$ will overestimate the actual log(O/H) abundance by a factor
of 0.2-0.5\,dex using a sample of 20 H~{\sc ii} regions in
  M101 with high-S/N spectra.
  Some other researches found the similar results, for example,
Bresolin et al. (2004, 2005), Garnett et al. (2004a,b),
Pilyugin (2006), Shi et al. (2005, 2006).
 A much larger dataset can help to understand better this effect 
and extending it to galaxies will also be of considerable interest.

To estimate abundances of galaxies, when $T_e$ and $R_{23}$ cannot be
used, some other metallicity-sensitive ``strong-line" ratios are very
useful, for example, 
[N~{\sc ii}]$\lambda $6583/H$\alpha $, 
[O~{\sc iii}]$\lambda $5007/[N~{\sc ii}]$\lambda $6583, 
[N~{\sc ii}]$\lambda $6583/[O~{\sc ii}]$\lambda $3727, 
[N~{\sc ii}]$\lambda $6583/[S~{\sc ii}]$\lambda\lambda $6717,6731, 
[S~{\sc ii}]$\lambda\lambda $6717,6731/H$\alpha $, and 
[O~{\sc iii}]$\lambda\lambda $4959,5007/H$\beta $ 
(Liang et al. 2006; Nagao et al. 2006; P\'{e}rez-Montero \& D\'{i}az 2005; 
Pettini \& Pagel 2004, hereafter
PP04; Denicol$\acute{o}$ et al. 2002, hereafter D02; Kewley \& Dopita
2002, hereafter KD02). Even when $R_{23}$ is available, some of these
line-ratios are useful to overcome the well-known problem that
  the $R_{23}$ vs. 12 + $\log$O/H relation is double-valued
   and further information is required to break
  this degeneracy. Alternative methods for breaking the degeneracy
  when a well established luminosity-metallicity relation exists, or
  when the likelihood of each branch can be calculated, which has been
  discussed by Lamareille et al. (2006).

Strong-line abundance indicators are typically calibrated in
  one of two ways: (1) using the samples of galaxies with direct
($T_e$-based) abundances (e.g. PP04; Pagel et al. 1979 etc.); (2)
using photoionization models (e.g. McGaugh 1991; Tremonti et al. 2004
etc.). Since the $T_e$-method is generally thought to be the
most accurate method for metallicity estimation, we will take
  oxygen abundances derived using this method as our baseline. We
select a large sample of 531 galaxies from the SDSS-DR4 with their
[O~{\sc iii}]$\lambda$4363 emission line detected at a S/N ratio
greater than 5$\sigma$, and 164 associated galaxies and H~{\sc ii} regions
from literature.  We compare their $T_e$-based O/H abundances
with the Bayesian estimates provided by the MPA/JHU group
(shown as log(O/H)$_{\rm Bay}$), which were
obtained by fitting multi-emission lines using the photoionization
models of Charlot et al. (2006), and with those derived from some
strong-line ratios given in previous studies, including $R_{23}$, $P$,
N2 and O3N2 methods etc. (Kobulnicky et al. 1999; Pilyugin 2001;
PP04). We also compare the observational results with the photoionization
model results of KD02 for the relations of O/H vs. $R_{23}$, N2, O3N2
and S2 indices.

Specially, we study the abundance calibrations of
  strong-line ratios on the basis of their $T_e$-based metallicities.
When we compare the observational relations of O/H vs. strong-line
ratios with the photoionization model results of KD02, we
find that only the N2, O3N2, S2 indices are useful to
estimate abundances for galaxies with low metallicity,
12+log(O/H)$_{T_e}$$<$8.5, while other line ratios, such as
[N~{\sc ii}]/[O~{\sc ii}], [N~{\sc ii}]/[S~{\sc ii}] are
[O~{\sc iii}]/H$\beta$ are not good indicators for this metallicity
range due to their insensitivity to metallicities there, except in
the extremely metal-poor environments (e.g. 12+log(O/H)$<$7.5 or 7.0)
(Stasi\'{n}ska 2002; KD02).
 $R_{23}$ is a useful indicator of metallicity
for the low metallicity region with 12+log(O/H)$<$7.9.
Thus, we will only calibrate the relationships of the
$R_{23}$, N2, O3N2 and S2 indices to O/H abundances from the
observational data in this study.  Moreover, for the N2 index, we
follow Pilyugin (2000;2001a,b) to add  the excitation parameter $P$
(=[O~{\sc iii}]/([O~{\sc ii}]+[O~{\sc iii}]))
 to separate the sample galaxies into three sub-samples in the
calibrations.
These calibrations can be the extension of the metal-rich
region studied by Liang et al. (2006) to the low metallicity region.

This paper is organized as follows. The sample selection criteria are
described in Sect.\ 2. The determinations of the oxygen abundances
from $T_e$ are presented in Sect.\ 3.  In Sect.\ 4, we present the
comparisons between the (O/H)$_{T_e}$ and the (O/H)$_{\rm Bay}$, as
well as those abundances derived from other strong-line
  relations.  Sect.\ 5 shows the comparison of the observational data
with the photoionization models of KD02. In Sect.6, we re-derive
analytical calibrations between O/H and $R_{23}$, N2, O3N2, S2 indices, as well
the two-parameter calibrations for the N2 index with $P$-parameter
included.  The conclusions are given in Sect.\ 7.

\section{Observational data}

\subsection{The SDSS-DR4 data}

The SDSS-DR4 (Adelman-McCarthy et al 2006) provides
spectra in the wavelength range $\sim$ 3800--9200\AA\ for $>$ 500,000
galaxies over 4783 square
degrees\footnote{http://www.sdss.org/dr4/}. The MPA-JHU
  collaboration has in addition made measurements of emission-line
fluxes and some derived physical parameters for a sample of 520,082
unique galaxies at the MPA SDSS
website\footnote{http://www.mpa-garching.mpg.de/SDSS/}.
 Therefore, we will call the working sample selected in this study
as the ``MPA/JHU sample" hereafter.
We
select the ``star-forming galaxies'' with metallicity measurements,
which have been identified following the selection criteria of the
traditional line diagnostic diagram [N~{\sc ii}]/H$\alpha$ vs. [O~{\sc
  iii}]/H$\beta$ (Baldwin et al. 1981; Veilleux \& Osterbrock 1987;
Kewley et al. 2001; Kauffmann et al. 2003).  The fluxes of
emission-lines have been measured from the stellar-feature subtracted
spectra with the spectral population synthesis code of Bruzual \&
Charlot (2003) (Brinchmann et al. 2004; Tremonti et al. 2004).

We select the galaxies with redshifts 0.03$<z<$0.25 to ensure to cover
from [O~{\sc ii}] to H$\alpha$ and [S~{\sc ii}] emission lines.
Tremonti et al. (2004) also discussed the weak effect of aperture on
estimated metallicities of the sample galaxies with $0.03 < z < 0.25$
and this was further discussed by Kewley et al. (2005), but
  for the present study the aperture effects are unimportant.

In this study, we use the $T_e$-method to derive O/H abundances of the
sample galaxies (see Sect.\,3 for details), therefore, we select the
samples with [O~{\sc iii}]$\lambda$4363 emission-line detected at a
S/N ratio greater than 5$\sigma$.  To be consistent with Liang et al.
(2006) and Tremonti et al. (2004), we also consider the objects having
been measured fluxes of [O~{\sc ii}]$\lambda\lambda$3726,3729, [O~{\sc
  iii}]$\lambda$5007, H$\beta$, H$\alpha$, [N~{\sc ii}]$\lambda6583$,
[S~{\sc ii}]$\lambda\lambda6717,6731$ emission lines, and the S/N
ratio of H$\beta$, H$\alpha$, [N~{\sc ii}], [S~{\sc ii}] are larger
than 5$\sigma$. The final sample consists of 531
  galaxies. These have fluxes in
[O~{\sc iii}]$\lambda$4363 greater than
5.3$\times$10$^{-17}$ ergs s$^{-1}$ cm$^{-2}$, with a mean value
of 21.27 $\times$10$^{-17}$ ergs s$^{-1}$ cm$^{-2}$.

The fluxes of the emission lines are corrected for dust extinction,
which are estimated using the Balmer line ratio H$\alpha$/H$\beta$:
assuming case B recombination, with a density of 100 cm$^{-3}$ and a
temperature of $10^4$ K, and the intrinsic ratio of H$\alpha$/H$\beta$
is 2.86 (Osterbrock 1989), with the relation of
$(\frac{I_{H\alpha}}{I_{H\beta}})_{obs}$=
$(\frac{I_{H\alpha0}}{I_{H\beta0}})_{intr}$10$^{-c(f(H\alpha)-
  f(H\beta))}$.  Using the average interstellar extinction law given
by Osterbrock(1989), we have $f$(H$\alpha$)- $f$(H$\beta$) = -0.37.
For the 56 data points with $c < 0$, we assume they have $c$ = 0 since
their intrinsic H$\alpha$/H$\beta$ may be lower than 2.86  if their
electron temperature is high (Osterbrock 1989, p.80).

 Nearly all previous empirical oxygen abundance calibrations
have been derived from  individual H~{\sc ii} regions since it is
much easier to detect [O~{\sc iii}]4363 that way.  In contrast the
SDSS data samples the inner few kpc of most galaxies. One question
is whether the global spectrum from a mixture of different H~{\sc
ii} regions will yield meaningful average abundances of the
galaxies or not. Kobulnicky et al. (1999), Moustakas \& Kennicutt
(2006) and Pilyugin et al. (2004) concluded that the spatially
unresolved emission-line spectra can reliably indicate the
chemical properties of distant star-forming galaxies. However, as
Kobulnicky et al. (1999) mentioned, the standard nebular chemical
abundance measurement methods may be subject to small systematic
errors when the observed volume includes a mixture of gas with
diverse temperatures, ionization parameters, and metallicities.
For the low-mass, metal-poor galaxies, as those we are studying in
this work, standard chemical analyses using global spectra will
overestimate the electron temperatures $T_e$ due to the
non-uniform $T_e$  and large variations in the ionization
parameter since the global spectra are biased to the objects with
stronger emission lines.  As a result, the oxygen abundances
derived from $T_e$  will be  underestimated, i.e. about
$<$0.1\,dex in log(O/H) (for more massive metal-rich galaxies like
local spiral galaxies, it is about $\pm$0.2\,dex discrepancy).
However, since this bias is small, and not well constrained for
our dataset, we do not attempt to correct for it.

\subsection{The metal-poor galaxies from literature}

In addition to the MPA/JHU samples, we also collect 164 low-metallicity
samples including some blue compact galaxies (BCDs) and
H~{\sc ii} regions from literature,
which are taken from Izotov et al. (1994, 1996, 1997a, 1997b, 1998a,
1998b, 1999a, 2001a, 2001b, 2004a, 2004b), van Zee (2000),
Kniazev et al. (2000), Vilchez et al. (2003), Guseva et al.
(2003a,b,c), Melbourne et al. (2004), and Lee et al. (2004).
 There are 110 H~{\sc ii} regions and 54 galaxies in this sample.

We re-estimate their O/H abundances by using the electronic
temperatures method given in Sect.\,3.  Their metallicities are
7.1$<$12+log(O/H)$<$8.4, more metal-poor than the SDSS galaxies
generally.

\begin{figure}
\centering
\includegraphics[width=7.8cm]{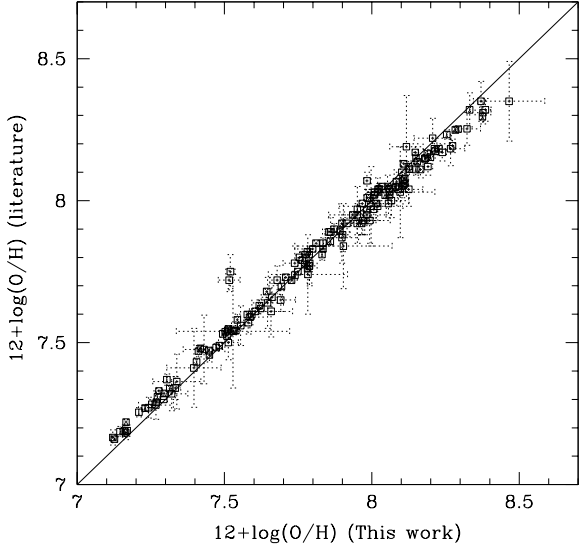} 
\caption{Comparison between the $T_e$-based oxygen abundances
(this work) and those from literature. } 
\label{fig1}
\end{figure}

To check if there is systematic difference between our estimates and
the values given in the previous studies, we compare their $T_e$-based oxygen
abundances obtained by us with those $T_e$-based given in the original reference
in Fig.~\ref{fig1}, which shows that they are very consistent, and the
very slight difference may come from different atomic data.

\section{Abundance determination from $T_e$}
\label{sect.3}

A two-zone model for the temperature structure within the H\,{\sc ii}
region was adopted. In this model, $T_e$([O~{\sc iii}]) is taken to
represent the temperature for high-ionization species such as
$O^{++}$, while $T_e$([O~{\sc ii}]) is used for low-ionization species
such as $O^{+}$.  The general method is firstly to derive $t_3$
(=10$^{-4}T_e$([O~{\sc iii}])) from the emission-line ratio of [O~{\sc
  iii}]4959,5007/[O~{\sc iii}]4363, and then to estimate $t_2$
(=10$^{-4}T_e$([O~{\sc ii}])) from an analytical relation between
$t_2$ and $t_3$ inferred from photoionization calculations.

Izotov et al. (2005) have recently published a set of equations for
the determination of the oxygen abundances in H {\sc ii} regions for a
five-level atom. They used the atomic data from the references listed
in Stasi\'{n}ska (2005).  According to those authors, the electron
temperature $t_3$ (in units of $10^4$ K), and the ionic abundances
$O^{++}/H^+$ and $O^{+}/H^+$ are estimated as follows:
\begin{equation}
t_3 = \frac{1.432}{{\rm log}((\lambda4959 + \lambda5007)/\lambda4363) -
{\rm log}C_T},
\end{equation}
where
\begin{equation}
C_T = (8.44 - 1.09t_3 + 0.5t_3^2 - 0.08t_3^3)\frac{1 + 0.0004x_3}{1+0.044x_3},
\end{equation}
with $x_3 = 10^{-4}n_et_3^{-1/2}$, and $n_e$ is the electron density in cm$^{-3}$.
And
\begin{eqnarray}
12 + {\rm log}(O^{++}/H^+) =
{\rm log}(I_{{[O}{III]\lambda4959+\lambda5007}}/I_{H\beta}) + \nonumber \\
                 6.200 + \frac{1.251}{t_3} - 0.55{\rm log}t_3 - 0.014t_3,
\end{eqnarray}
\begin{eqnarray}
12 + {\rm log}(O^{+}/H^+) =
{\rm log}(I_{{[O}{III]\lambda3726+\lambda3729}}/I_{H\beta}) +  \nonumber \\
      5.961 + \frac{1.676}{t_2} - 0.40{\rm log} t_2 - 0.034 t_2 +  \nonumber \\
      {\rm log}(1 + 1.35x_2),
\end{eqnarray}
with $x_2 = 10^{-4}n_et_2^{-1/2}$,
and $n_e$ is the electron density in cm$^{-3}$.

   The total oxygen abundances are then derived from the following
equation:
\begin{equation}
\frac{O}{H} = \frac{O^+}{H^+} + \frac{O^{++}}{H^+}.
\end{equation}

The electron temperature $t_2$ (in units of $10^4$ K) of the
low-ionization zone is usually determined from $t_3$ following an
equation derived by fitting H {\sc ii} region models. Several versions
of this relation of $t_2$ vs. $t_3$ have been proposed. We use the one
of Garnett (1992), which has been widely used:
\begin{equation}
t_2 = 0.7t_3 + 0.3.
\end{equation}

The electron densities in the ionized gas of the galaxies are
calculated from the line ratios [S~{\sc ii}]$\lambda6717$/[S~{\sc
  ii}]$\lambda6731$ by using the five-level statistical equilibrium
model in the task {\sc temden} contained in the {\sc iraf/stsdas}
package (de Robertis, Dufour \& Hunt 1987; Shaw \& Dufour 1995),
which uses the latest atomic data.  A reasonable upper limit of
the ratios is 1.431 (Osterbrock 1989).  However, we find that 118
out of the 531 SDSS galaxies have [S {\sc ii}] line ratios larger
than 1.431.  For 71 of the 118 galaxies the difference between the
measured ratio and 1.431 was less than the error in the line
ratio.  So it is very possible that
  most of the high
[S~{\sc ii}]$\lambda6717$/[S~{\sc ii}]$\lambda6731$ ratios come from errors.
Therefore, we adopt 1.431 for these galaxies. Indeed,
this approximation would not affect much the ionic abundances.  The
reasons are: the term containing $x_3$ is never important and
can be omitted in calculating $C_T$ since $n_e$ is always smaller
than 10$^3$ cm$^{-3}$; and $x_2$ is also insignificant since it is
generally less than 0.1 with $n_e$ $<$ 10$^3$ cm$^{-3}$.

Oxygen abundances of all the 695 samples (531 from SDSS-DR4 and 164
from literature) are estimated using the relations outlined
  above. The whole sample show a metallicity range of 7.1 $<$ 12 +
log(O/H)$_{T_e}$ $<$ 8.5 mostly. Most of the SDSS-DR4 galaxies
lie in the more metal-rich region, i.e., 7.6 $<$ 12 +
log(O/H)$_{T_e}$ $<$ 8.5, and only 63 of them have lower metallicities
than 12 + log(O/H)$_{T_e}$ = 8.0.  Most of the samples from
literature have lower metallicities, i.e., 12 + log(O/H) $<$ 8.0. The
typical uncertainty of the estimates is about 0.044\,dex in
12+log(O/H), and about 500\,K in $T_e$([O~{\sc iii}]).  All the data
for the SDSS galaxies are given in Table\,1.

\section{Comparisons between the $T_e$ and strong-line oxygen abundances}

\subsection{Comparison with the Bayesian metallicities obtained by the
MPA/JHU group}

It will be very interesting to compare the $T_e$-based O/H abundances
with the Bayesian oxygen abundances provided by the MPA/JHU
group for the sample galaxies.
The MPA/JHU group used photoionization model (Charlot et al. 2006)
to fit simultaneously most prominent emission lines.
 Based on the Bayesian technique, they
calculated the likelihood distribution of the
metallicity of each galaxy in the sample by comparing with a large
library of models ($\sim$2$\times$10$^5$) corresponding to different
assumptions about the effective gas
parameters, and then they adopted the median of the distribution
as the best estimate of the galaxy metallicity
(Tremonti et al. 2004; Brinchmann et al. 2004).

\begin{figure}
\centering
\includegraphics[bb=73 278 245 635,width=7.5cm,clip]{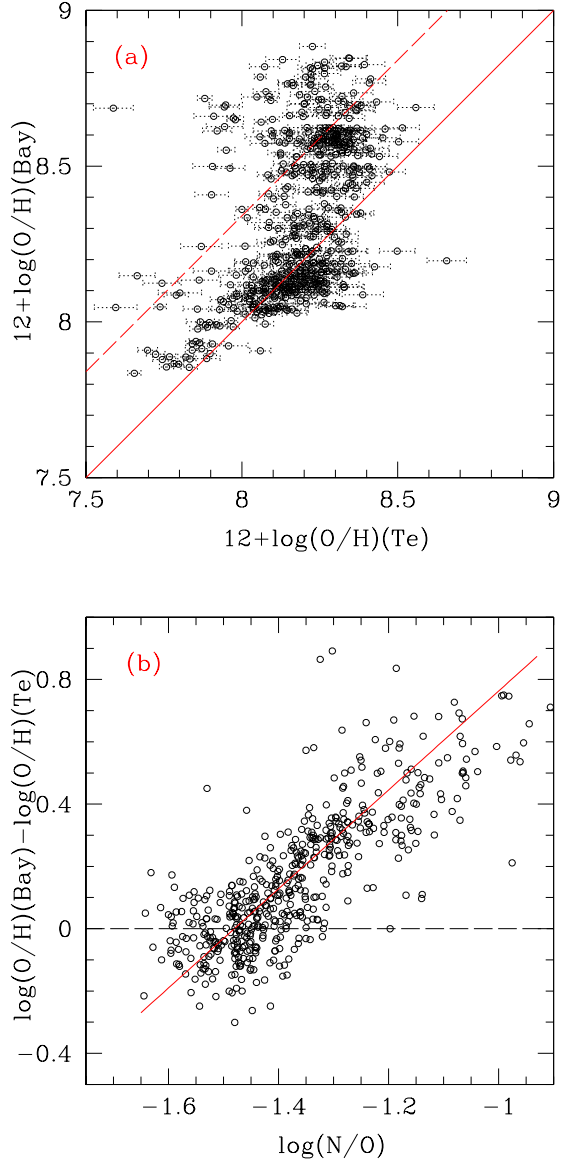}
\caption {{\bf (a)} Comparison between the electron temperature-based oxygen abundances
and those Bayesian estimates obtained by the MPA/JHU group.
The solid line is the equal-value line, and the
long-dashed line represents a +0.34\,dex discrepancy from the
equal-value line to show the overestimates of the oxygen
abundances by using the model of Charlot et al. (2006)
for almost half of the sample galaxies.
{\bf (b)} The offset between the log(O/H)$_{\rm Bay}$ and the log(O/H)$_{T_e}$ as a function of the
log(N/O) abundance ratios for the MPA/JHU sample galaxies. The solid line is
the linear least squares fit for the relations, given as Eq.~(\ref{offset}).
}
\label{fig2}
\end{figure}

Very surprisingly, Fig.~\ref{fig2}a shows that, for almost half of the
sample galaxies ($\sim$227), their metallicities are overestimated by
a factor of about 0.34\,dex on average by using the model of Charlot et al.
(2006).  There is no obviously  monotonic trend
between the increasing discrepancy and the weakening [O~{\sc iii}]4363
line or the decreasing [O~{\sc iii}]4363/H$\beta$ ratio.  What is the
reason for such obvious difference?

Indeed, Charlot's model was based on Charlot \& Longhetti (2001, CL01
hereafter), which is a model for computing consistently the line
emission and continuum from galaxies based on a combination of recent
population synthesis and photoionization codes of CLOUDY. What they
did is to use a sample of 92 local galaxies as the observed
constraints to determine the best model parameters, including
metallicity Z etc.  In their calibrations, they used the emission-line
luminosities of H$\alpha $, H$\beta $, [O~{\sc ii}]$\lambda $3727,
[O~{\sc iii}]$\lambda $5007, [N~{\sc ii}]$\lambda $6583 and [S~{\sc
  ii}]$\lambda\lambda $6717,6731 emission lines. The main
  difference from this method and the other strong line methods
  discussed below is that all the emission lines are used
  simultaneously so that any clear change in the abundance ratio $X/O$
  for any element $X$ from that used in the CL01 models has the
  potential of causing offsets. In particular, the clear offset
  between (O/H)$_{Bay}$ and (O/H)$_{T_e}$ is possibly related to how
  secondary nitrogen enrichment is treated in the CL01 models.

 Fig.~\ref{fig2}b shows that the ``offset" between log(O/H)$_{Bay}$
and log(O/H)$_{T_e}$ (the former minus the latter)
 strongly correlates with the
log(N/O) abundance ratio, which can be shown as a linear least squares fit:
\begin{equation}
\label{offset}
 \rm offset = 1.589 \times log(N/O) + 2.352,
\end{equation}
with a rms of 0.176\,dex.
The log(N/O) abundance
ratios are obtained from the electron temperature in the [N~{\sc ii}]
emission region (is equal to the $t_2$ given in Sect.~\ref{sect.3})
  and the flux ratio of [N~{\sc ii}]$\lambda\lambda
$6548,6584 to [O~{\sc ii}]$\lambda $3727 following the method given in
Thurston et al. (1996).
This means that the overestimates of
CL01 models for the O/H abundances may be related to the onset of
secondary N enrichment. There is a considerable spread in the
onset of N enrichment and therefore in this transition region
the assumption of the CL01 models of a single N enrichment trend is too
simplistic. The simple way to avoid this problem might be to exclude
[N~{\sc ii}] from their fitting procedure.
In this
transition region the relations between the log(N/O) and the log(O/H)
of galaxies are not monotonic, not as the
almost constant trend in the low-metallicity region
or the increasing trend in the high-metallicity region
(see fig.8 of Liang et al. 2006).

\subsection{$R_{23}$ method}

Pagel et al.(1979) first proposed the empirical abundance indicator
$R_{23}$ for metal-rich galaxies. Skillman et al. (1989)
further suggested that this indicator also can be used for
metal-poor galaxies. The relationship between O/H and $R_{23}$
  has been extensively discussed in literature (see Tremonti et
al. (2004) and references therein).  Different calibrations will
result in slightly different oxygen abundances even for the same
branch (see fig.\ 3 of Tremonti et al. 2004 for comparison). A
  few calibrations have attempted to improve the calibration by
  introducing an empirical estimator of the ionization parameter as a
  second parameter (e.g. McGaugh 1991; Charlot \& Longhetti 2001;
KD02 etc.).  At present, one of the popular formulas is
the calibration given by Kobulnicky et al.(1999), which includes two
analytical formulas, for the metal-rich and metal-poor branches
respectively, and is derived from the photoionization model of McGaugh
(1991).

As we mentioned in the introduction, $R_{23}$ may overestimate the
actual O/H abundances by a factor of $\sim$0.2-0.5\,dex (Kennicutt et
al. 2003 etc.). On the basis of this large sample of 695 galaxies, here we
compare their oxygen abundances derived from $R_{23}$ and $T_e$
methods. Figs.~\ref{fig3}a,3b show this comparison, where we use the
calibration formula of Kobulnicky et al. (1999) to derive the
12+log(O/H)$_{R_{23}}$ for the sample galaxies.  In Fig.~\ref{fig3}a,
for the galaxies with 12+log(O/H)$_{T_e}$$<$7.95, we use their
calibration formula for metal-poor galaxies; and for the galaxies with
12+log(O/H)$_{T_e}$$\geq$8.2, we use their formulas for metal-rich
branch.  These two metallicity limits are defined following the
$R_{23}-P$ method in Sect.\ 4.3 and some other related studies (Pilyugin
2000, 2001a,b; Shi et al.\ 2006).  Fig.~\ref{fig3}a shows that
$R_{23}$ method will overestimate the O/H abundances by a factor of
$\sim$0.20\,dex, which is analogous to what was found by Kennicutt et
al.  (2003) and Shi et al. (2005, 2006).  Detailed discussions about
this discrepancy can be found in Kennicutt et al. (2003).  For the
sample galaxies with 8.0$<$12+log(O/H)$_{T_e}$$<$8.2, which are in the
turn-over region, the calibration of Kobulnicky et al. (1999) of
metal-poor branch gives (O/H)$_{R_{23}}$ in good agreement
  with (O/H)$_{T_e}$.  These galaxies are omitted when we discuss the
discrepancy between the $T_e$- and $R_{23}$-based metallicities,
which is acceptable.  Fig.~\ref{fig3}a does not show any obvious
offset between the H~{\sc ii} regions and galaxies.
P{\'e}rez-Montero \& D{\' i}az (2005) showed that the giant
extragalactic H~{\sc ii} regions (GEHRs) have more dispersion than
the H~{\sc ii} galaxies and H~{\sc ii} regions in the Galaxy and
the Magellanic Clouds (their figs.1,3,7) since the GEHRs may have
different ionizaiton parameters and stellar effective
temperatures.

To illustrate clearly the uncertainties due to the double-valuedness
of the $R_{23}$-O/H relation, we used both of the calibration
formulas for metal-poor and metal-rich branches to calculate the
12+log(O/H)$_{R_{23}}$ abundances for the galaxies having
7.9$<$12+log(O/H)$_{T_e}$$<$8.4.  Fig.~\ref{fig3}b gives the results
as the points between the two vertical dashed lines, which shows that
these two different formulas will result in quite different
metallicities, and the discrepancy may be up to $\sim$ 0.4 dex.
Therefore, one should be very careful when using $R_{23}$-method to
derive the metallicities of the galaxies having metallicities in the
turn-over region.

\subsection{The $R_{23}-P$ method}

Pilyugin (2000, 2001a,b) suggested the $P$ method to estimate oxygen
abundances of galaxies, in which the oxygen abundance can be
derived from two parameters, $R_{23}$ and $P$
(=[O~{\sc iii}]/([O~{\sc ii}]+[O~{\sc iii}])).
 Therefore, we call this method as
the ``$R_{23}-P$ method".  They derived the O/H=$f(R_{23},P)$
formulas from a sample of metal-poor H~{\sc ii} regions with
7.1$<$12+log(O/H)$_{T_e}$$<$7.95 and a sample of moderately metal-rich
 H~{\sc ii} regions with 8.2$<$12+log(O/H)$_{T_e}$$<$8.7.  The best fitting
relation for metal-poor objects was given as Eq.(4) in Pilyugin
(2000), and that for moderately metal-rich objects was given as
Eq.(8) in Pilyugin (2001a).

Fig.~\ref{fig3}c shows that there is slight difference of
  $\sim$0.06\,dex between the $(O/H)_{T_e}$ and the $(O/H)_{R_{23}-P}$
abundances for the samples.  Indeed, the good
  agreement is not surprising since Pilyugin (2000, 2001a,b) derived
  the $R_{23}-P$ method formulas by using a subset of the sample we use
  here (Izotov et al. 1994, 1997a and Izotov \& Thuan 1998b,1999b). The
  minor offset is likely caused by two reasons: (1) our sample is
much larger than that used by Pilyugin which allows us to determine
the relations with higher precision; (2) the different $T_e$-formulas
and/or atomic data used in calculations.
Moreover, for the low excitation samples in
low-metallicity branch with
$P<$0.75, the $R_{23}-P$ method may overpredict their
oxygen abundances by a factor $\sim$0.1\,dex, which
is consistent with the 0.1-0.2\,dex overestimates found
by van Zee \& Haynes (2006)
for the low excitation H {\sc ii} regions in dwarf irregular galaxies.

Pilyugin \& Thuran (2005) have renewed the $R_{23}-P$ calibrations
by including several improvements, such as enlarging the sample by
1 order of magnitude etc. We have used the new calibrations to
re-calculate the (O/H)$_{R_{23}-P}$ abundances for our sample
galaxies, and found the overestimate factor by the $R_{23}-P$
method decreases to be $\sim$0.025\,dex.

 Because there is no observational samples with
 7.95$<$ 12+log(O/H)$<$ 8.2 in Pilyugin's calibration for $R_{23}-P$ method,
 the $R_{23}-P$ method cannot provide information for the galaxies within
 this oxygen abundance range. To understand more about this,
  we calculate oxygen
  abundances for galaxies with 12+log(O/H)$_{T_e}$=7.9--8.4 using the two
  formulas for the metal-rich and metal-poor branches. The results are
  indicated as the points between the two vertical dashed lines in
  Fig.~\ref{fig3}d.  Similarly to what we saw for $R_{23}$ the two
  branch calibrations give quite different metallicities with a difference of up
  to $\sim$0.4\,dex.  Thus one must be very careful when applying the
  $R_{23}-P$ method to samples of galaxies with metallicities in the turn-over
  region.

\subsection{$N$2 method}

The [N {\sc ii}]/H$\alpha$ emission-line ratio can also be used to
estimate metallicities of galaxies. Nitrogen is mainly synthesized in
intermediate- and low-mass stars. The [N {\sc ii}]/H$\alpha$ ratio
 will therefore become stronger with on-going star formation
and evolution of galaxies, until very high metallicities are
  reached, i.e.  12+log(O/H)$>$9.0, where [N {\sc ii}] starts to
become weaker due to the very low electron temperature caused from
strong cooling by metal ions. It is worth pointing out that
  this ratio gives an indirect measurement of the oxygen abundance
  however, since oxygen lines are not used in the method.

Following the earlier work by Storchi-Bergmann, Calzetti \&
Kinney(1994) and Raimann et al.(2000), D02 and PP04 suggested
calibration relations for O/H vs. N2 index from 236 and 137 galaxies
respectively.  D02 combined a sample of 128 metal-rich galaxies and
108 metal-poor galaxies with [O~{\sc iii}]4363 detected
to derive their calibration.  The way
  they estimated the metallicities of their galaxy sample was to use
  the $T_e$-method for the metal-poor galaxies and the $R_{23}$ or
  $S_{23}$ (=log(([S~{\sc ii}]6717,6731+[S~{\sc
    iii}]9096,9531))/H$\beta$) method for the metal-rich ones. PP04
extracted the metal-poor objects from the work of D02 and updated the
abundances for some of them with recent measurements by
Kennicutt et al. (2003) for H~{\sc ii} regions in M101 and
by Skillman et al. (2003) for dwarf irregular galaxies in the Sculptor Group.
  To be consistent with the
$T_e$-method, here we use the calibration formula given by PP04 to
derive the metallicities 12+log(O/H)$_{N2}$ for our 695 sample
galaxies.  Fig.~\ref{fig3}e gives the comparison between the
12+log(O/H)$_{T_e}$ and 12+log(O/H)$_{N2}$, which shows that the two
estimates are consistent but with some scatter.  The scatter may
mostly come from the different ionization parameters of the galaxies,
and the different samples (see Sect.\ref{sec6.5}).

One of the advantages of using the N2 index for estimating metallicity
is that the $N$2 indicator can break the degeneracy when the
  $R_{23}$-(O/H) and ($R_{23},P$) -(O/H) relations are double-valued.  Also,
the N2 index is not affected by dust extinction since
  H$\alpha$ and [N~{\sc ii}] are so closely spaced in wavelength.
  Furthermore near infrared spectrographs can measure these two lines
for galaxies out to high redshifts, where equivalently to early
  epochs in the evolution of galaxies (see more discussion in Liang et al.
2006).

\subsection{$O$3$N$2 method}

\begin{figure}
\centering
\includegraphics[width=7.8cm]{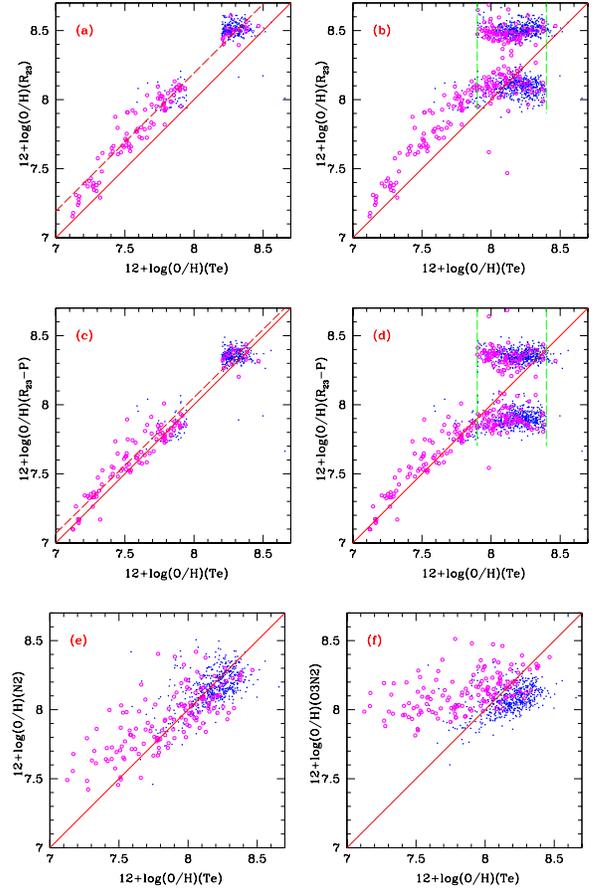}   
 \caption {Comparison between the $T_e$-based abundances with those
  derived from other strong-line calibrations: {\bf (a,b).} $R_{23}$
  method; {\bf (c,d).} $R_{23}-P$ method; {\bf (e).} $N2$ method; {\bf (f).}
  $O3N2$ method (see text).  The small blue points represent the SDSS
  sample, and the larger magenta points (open circles) represent the
  sample compiled from literature.  The solid lines are the
  equal-value lines, and the long-dashed lines in {\bf (a)} and {\bf
    (c)} are the linear least-squares fits to the data points, which
  show a discrepancy of $\sim$0.20 dex and $\sim$0.06 dex from the
  equal-value lines, respectively.  The two pairs of vertical dashed
  lines in {\bf (b)} and {\bf (d)} show the range of
  7.9$<$12+log(O/H)$_{T_e}$$<$8.4 for the samples whose metallicities
  are estimated from strong-line ratios by using two calibrations for
  metal-rich and metal-poor branches.  }
\label{fig3}
\end{figure}

$O$3$N$2 = log\{([O~{\sc iii}]$\lambda5007$/H$\beta$)/([N~{\sc
  ii}]$\lambda6583$/H$\alpha$)\} is another indicator for
  metallicities of galaxies.  It was introduced by
Alloin et al.(1979) and further studied by PP04. PP04 presented one
such calibration by doing linear least squares fit for a sample of 65
(from the 137) galaxies with $-1 < O3N2<$ 1.9, and provided a
calibration formula. This calibration does not work for the cases of
$O$3$N$2 $> 1.9$ since the data points are so scattered there in their
study.  Fig.~\ref{fig3}f shows the comparison between the
12+log(O/H)$_{T_e}$ and the oxygen abundances derived from O3N2 method
using the calibration of PP04 for our 695 sample galaxies.  But we
extrapolate to use the PP04's calibration to the galaxies with
$O3N2$$>$1.9.

It is clear that this calibration will overestimate the O/H abundances
by a big factor, up to 1.0\,dex, for the metal-poor galaxies with
$O$3$N$2 $>$ 1.9 (about 12+log(O/H)$<$8.0).  The scatter of the data with
$-1<$$O$3$N$2 $<$ 1.9 are also large,
which may be due to the different ionization
parameters of the galaxies.

\section{Comparisons with photoionization models}

\begin{figure}
\centering
\includegraphics[width=7.8cm]{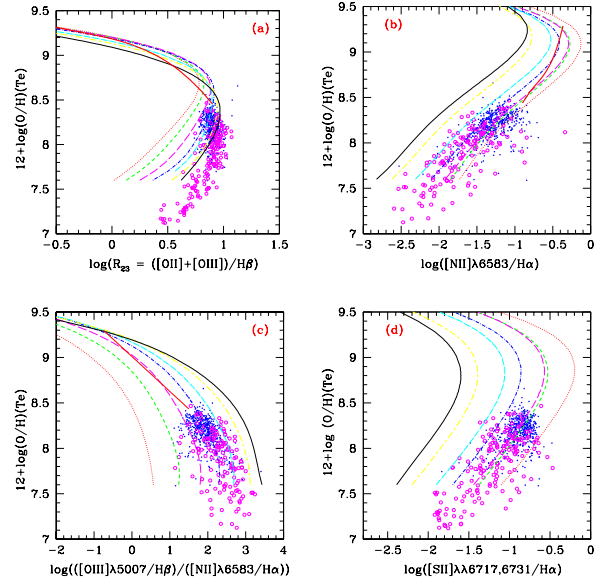}   
\caption{Comparison between the observational data
  and the photoionization models of KD02.  The symbols are the same as
  in Fig.~\ref{fig3}.  The seven lines represent the model results from
  KD02 with seven photoionization parameters from q = 5 $\times$
  $10^6$ cm s$^{-1}$ to q = 3 $\times$ $10^8$ cm s$^{-1}$, which
  increase orderly from the red dotted line to the black solid line.
  The solid thick line (in red) in Fig.(a) is the calibration derived
  by Tremonti et al. (2004) for the metal-rich galaxies,
  and those in Fig.(b) and Fig.(c) are
  the calibrations derived by Liang et al. (2006) for the metal-rich SDSS galaxies.
}
\label{fig4}
\end{figure}

KD02 has calculated the metallicity calibrations for strong-line
ratios on the basis of photoionization models. They used a combination
of stellar population synthesis and photoionization models to develop
a set of ionization parameters and abundance diagnostics based on
emission-line ratios from strong optical lines, with seven
ionization parameters q = 5 $\times$ $10^6$, 1 $\times$ $10^7$, 2
$\times$ $10^7$, 4 $\times$ $10^7$, 8 $\times$ $10^7$, 1.5 $\times$
$10^8$, and 3 $\times$ $10^8$ cm s$^{-1}$.  The ionization parameter
$q$ is defined on the inner boundary of the nebula, and can be
physically interpreted as the maximum velocity of an ionization front
that can be driven by the local radiation field.  It will be
interesting to compare the observational samples with these model
results. Fig.~\ref{fig4} shows the comparisons for the relations of
12+log(O/H) versus $R_{23}$, N2, O3N2 and S2 indices.
These relations are for the $T_e$-based oxygen abundances of the
metal-poor galaxies. Liang et al. (2006) has shown such relations
for the metal-rich SDSS galaxies based on $R_{23}$ abundances, and derived
the corresponding calibrations (see the thick solid
lines in Figs.~\ref{fig4}b,c).

Fig.~\ref{fig4}a shows that the significant discrepancy in the
theoretically expected $R_{23}$ sequence with respect to the observed
trend beginning from 12 + log(O/H)$_{T_e} = 8.0$ to lower
metallicities.  Nagao et al. (2006) also pointed out this.
The observed relation of $R_{23}$ - log(O/H) of the data points
does not show much scatter in contrast to the models of the lower-branch of
metallicity, which may mean the weak independence of  $R_{23}$ on
ionization parameters there.
Especially, the distribution of observational data is nearly vertical
in oxygen abundance range of $7.9 \sim 8.5$.  The physical reason for
the insensitivity of $R_{23}$ to metallicity in this metallicity region
is that the [O~{\sc iii}]/H$\beta$ ratio
(also [O~{\sc ii}]/H$\beta$) is independent of $T_e$,
hence of metallicity there (Stasinska 2002).
It proves again that the oxygen abundance cannot be accurately
measured by using $R_{23}$ in the metallicity turnover region,
 but $R_{23}$ indicator works well for the metal-poor galaxies
with 12 + log(O/H)$<$7.9.

Fig.~\ref{fig4}b shows the comparison between 12+log(O/H) and N2
index.  It shows that the observational data follow the general trend
of the model results well, namely, the N2 indices of galaxies increase
following the increasing metallicities, up to 12+log(O/H)=9.0 (by
models and Liang et al. 2006).  The observational data show some
scatter, which may be due to their different ionization parameters.
We also plot a thick solid line (the short one) in Fig.~\ref{fig4}b
to show the calibration relation derived by Liang et al.
(2006) with the metal-rich SDSS galaxies (8.5$<$12 + log(O/H)$<$9.3).
If we derive a calibration from the $T_e$-based oxygen
abundances of the metal-poor galaxies,
and extrapolate it to the metal-rich region, it will
result in relatively lower O/H abundance than directly using the
calibration from metal-rich galaxies.  Perhaps one reason for this
discrepancy is from the different methods to determine metallicities
of the metal-poor and metal-rich galaxies, i.e. the $T_e$-method
versus the $R_{23}$-method or photoionization models.  However,
Stasi\'{n}ska (2005) pointed out that, at high metallicity, the
derived O/H values from $T_e$ for model H~{\sc ii} regions deviate
strongly from the true ones, i.e., important deviations appear around
12 + log(O/H)=8.6, and may become huge as the metallicity increases
(see their fig.1a).

Fig.~\ref{fig4}c shows the comparison between 12+log(O/H) and the O3N2
index. The general trend of the observations and that of the
  models are similar, and show increasing O/H abundance with
decreasing O3N2 index.  However, the large scatter of
the data points in this plot may be due to the strong effect of ionization
parameter, and such effect is stronger in the
low-metallicity region (12+log(O/H)$<$8.5) than in the
high-metallicity region (12+log(O/H)$>$8.5).  Moreover, O3N2 is much
less dependent on metallicity in the low-metallicity region than in
the high-metallicity region, in particular, it is almost
independent of metallicity for the sample galaxies with metallicities
of 12+log(O/H)=7.1-7.7.  This is possibly the main reason that PP04 had to
limit their calibration to O3N2$<$1.9. We
also present the calibration for metal-rich SDSS galaxies obtained by
Liang et al. (2006) in Fig.~\ref{fig4}c, which is given as the thick solid
line (the short one) in the 8.5$<$12+log(O/H)$<$9.3 region.  Both the
calibrations for the metal-rich and metal-poor sample galaxies are
consistent with the general trend of the model predictions, but the
calibration for metal-poor galaxies corresponds to relatively higher
ionization parameter than the metal-rich galaxies.

Fig.~\ref{fig4}d shows the comparison between 12+log(O/H) and the S2
index. Both the observations and models clearly show an
  increasing S2 with increasing O/H. This increasing trend will
extend up to 12+log(O/H)$\sim$8.9-9.0.  S2 is affected more by the
ionization parameters than the N2 index is. The large scatter of
this relation may be caused by the different ionization parameters.
Most of the SDSS galaxies cluster around $-1.0<S2<-0.5$.
 Liang et al. (2006) did not derive a metallicity calibration for S2
for the metal-rich galaxies since the data points show obvious
turnover and double-valued distribution around 12+log(O/H)$\sim$8.9-9.0.

In summary, Figs.~\ref{fig4}a-d show that $R_{23}$ is not a good
metallicity indicator for the galaxies with 7.9$<$12+log(O/H)$<8.5$ due to
its independence on metallicity there, and is a good indicator
for lower metallicity region with 12+log(O/H)$<7.9$;
N2 is a useful indicator for estimating
metallicities of galaxies because it is strongly monotonicly correlated with
  metallicity (up to about 12+log(O/H)$\sim$9.0);
although O3N2 and S2 can be
used to estimate metallicities of galaxies, they are more sensitive to
ionization parameters than N2 is, and O3N2 is less dependent on
metallicity than the other two.  In
  addition, when we have the indicator of O3N2 and/or S2, normally we
also have N2.  Furthermore, if we can find a way to take into account
the ionization parameters of the sample galaxies in the calibrations,
it will provide more reliable metallicities (see Sects.~\ref{sec6.5},\ref{sec6.6}).

\section{Deriving oxygen abundance calibrations}

We have obtained the $T_e$-based O/H abundances for the 695 sample
galaxies and H~{\sc ii} region.
They have metallicities of $7.1
<$12+log(O/H)$_{T_e}$ $< 8.5$ mostly.  Then we use their (O/H)$_{T_e}$
and line ratios to re-derive the oxygen abundance calibrations for the
strong-line ratios N2, O3N2 indices, and add a S2 index.
We also derive the calibration for $R_{23}$ for the 12+log(O/H)$_{T_e}$ $< 7.9$
region. Moreover, we try to add an excitation parameter $P$
(=[O~{\sc iii}]/([O~{\sc ii}]+[O~{\sc iii}])) in the
calibrations to separate the sample galaxies to be three sub-samples,
which improves the N2 calibration, but does not improve
well the O3N2 and S2 calibrations.

\subsection{The $R_{23}$ index}

The $R_{23}$ index is useful for calibrating metallicities
of low metallicity galaxies with 12+log(O/H)$<$7.9 due to
its obvious correlation with log(O/H) and weak dependence on
ionization parameters there.
Fig.~\ref{fig5}a shows this correlation.
The linear least-squares fit for all the 120
data points with 12+log(O/H)$<$7.9 is:
\begin{equation}
\label{eqr23}
 12 + \rm log(O/H) = 6.486 + 1.401 \times log(R_{23}),
\end{equation}
 with $R_{23}$=([O~{\sc ii}]$\lambda$3727+
 [O~{\sc iii}]$\lambda$$\lambda$4959,5007)/H$\beta$. The rms of the data to
the fit is about 0.103\,dex. The appropriate range of this
calibration is about
 0.4$<$log($R_{23}$)$<$1.0 and 7.0$<$12+log(O/H)$<$7.9.
 The calibrations of Kobulnicky et al.(1999) are also presented
 in the figure for  comparison
 as the dashed lines for the cases of $y$ (=log([O~{\sc iii}]/[O~{\sc ii}]))
 equal to 0, 1, 1.5.  It shows that the $y$=1.0 and 1.5 can explain
 these observational data generally.

\subsection{The N2 index}
\label{sec6.2}

Since the data points show a relatively tight linear relationship
between $N$2 and 12+log(O/H), we obtain a linear least-squares
fit for this relation. We use the mean-value points in 11 bins
of metallicities to avoid too much weights on the high metallicity
range where most of the data are distributed.  The bin width is
$\Delta$[log(O/H)] = 0.1 for the galaxies in the range of $7.4
<$12+log(O/H) $< 8.5$, and $\Delta$[log(O/H)] = 0.3 for the range of
7.1 $<$ 12 + log(O/H)$_{T_e}$ $<$ 7.4 because of the small number of
sources in this range.  These bin widths are the same as
for the O3N2 and S2 indices discussed in the consequent two subsections.
The objects with higher uncertainty than
0.06\,dex on log(O/H) and log([N~{\sc ii}]/H$\alpha$) have been
excluded in calculating the mean values and calibrating, which leaves
584 data points. 19 galaxies from literature cannot be included in
deriving calibrations since they do not have [N~{\sc ii}]
emission-line measurements.  The derived analytical relation of a linear
least squares fit is:
\begin{equation}
\label{eqN2}
 12 + \rm log(O/H) = 9.263 + 0.836 \times N2,
\end{equation}
 with $N$2=log([N~{\sc ii}]$\lambda$6583/H$\alpha$),
shown as the solid line in Fig.~\ref{fig5}b.  The rms of the data to
the fit is 0.159\,dex. The two dashed lines show a spread of
2$\sigma$.  This calibration is valid in the range of $-2.5 < N2 <
-0.5$.  The calibrations of D02 ($12+\log{\rm (O/H)} = 9.12 + 0.73
\times N2$, the dotted line) and PP04 ($12 + \log {\rm (O/H)} = 8.90 +
0.57 \times N2$, the long-dashed line) are also plotted for comparison.
It shows that D02's calibration is
similar to ours, but the one of PP04 shows a slightly lower slope and the
difference is more obvious for the low metallicity region with
12+log(O/H)$<$8.0. The reason may be that PP04's sample
 does not extend to much lower metallicity with 12+log(O/H)$<$7.7,
and most of their samples have metallicities of
7.7$<$12+log(O/H)$<$8.5, while the fit for their these samples
results in a bit lower slope than for ours (see fig.1 of D02 and fig.1 of PP04).

\subsection{The O3N2 index}

Fig.~\ref{fig5}c shows our sample galaxies for their (O/H)$_{T_e}$
abundances versus O3N2 indices.  Although they
are quite scattered, there still exists an obvious trend, i.e., O/H
increases following the decreasing O3N2 index.  We try to derive a
calibration of O3N2 index for O/H abundances from these sample
galaxies.  The objects with an higher uncertainty than 0.06\,dex on
log(O/H) and O3N2 are excluded, which leaves 585 data points.
 19 objects from literature
without their [N~{\sc ii}] emission-line measurements are excluded.
The mean-value points within 11 bins are obtained for the left sample
galaxies. The bin widths are the same as for the N2 calibration
(Sect.~\ref{sec6.2}).
A two-order polynomial fit is obtained by fitting the mean-value points:
\begin{equation}
\label{eqO3N2}
 12 + \rm log(O/H) = 8.203 + 0.630 \times O3N2 - 0.327  \times O3N2^2,
\end{equation}
 with O3N2 = log[([O {\sc iii}]$\lambda$5007/H$\beta$)/([N {\sc ii}]$\lambda$6583/H$\alpha$)],
shown as the solid line in Fig.~\ref{fig5}c. The rms of the data to
the fit is 0.199\,dex. This fit is valid for 1.4$<$O3N2$<$3.
The two dashed lines show 2$\sigma$ discrepancy.
The calibration of PP04 ($12+\rm {log(O/H)} = 8.73 - 0.32 \times O3N2$)
has been plotted as the long-dashed line, which is only
valid for 1$<$O3N2$<$1.9. The calibration of PP04 shows a shallower
slope than ours, which comes from their much smaller sample (65
objects) and the much narrower O3N2 range.

\subsection{The S2 index}

S2 = log([S~{\sc ii}]$\lambda\lambda$6717,6731/H$\alpha$) is also a
metallicity-sensitive indicator.  Although it is less useful than N2
for metallicity calibration due to the stronger dependence on
ionization parameter, it is still interesting to derive a calibration
from S2 since it is dependent on metallicity monotonicly in the
studied metallicity range.  This relation has the advantage of not
being strongly dependent on reddening corrections.  Although the
sample data is much scatter, a linear relation exists obviously between O/H
and S2.

\begin{figure*}
\centering
\includegraphics[bb=69 349 533 706,width=15.2cm,clip]{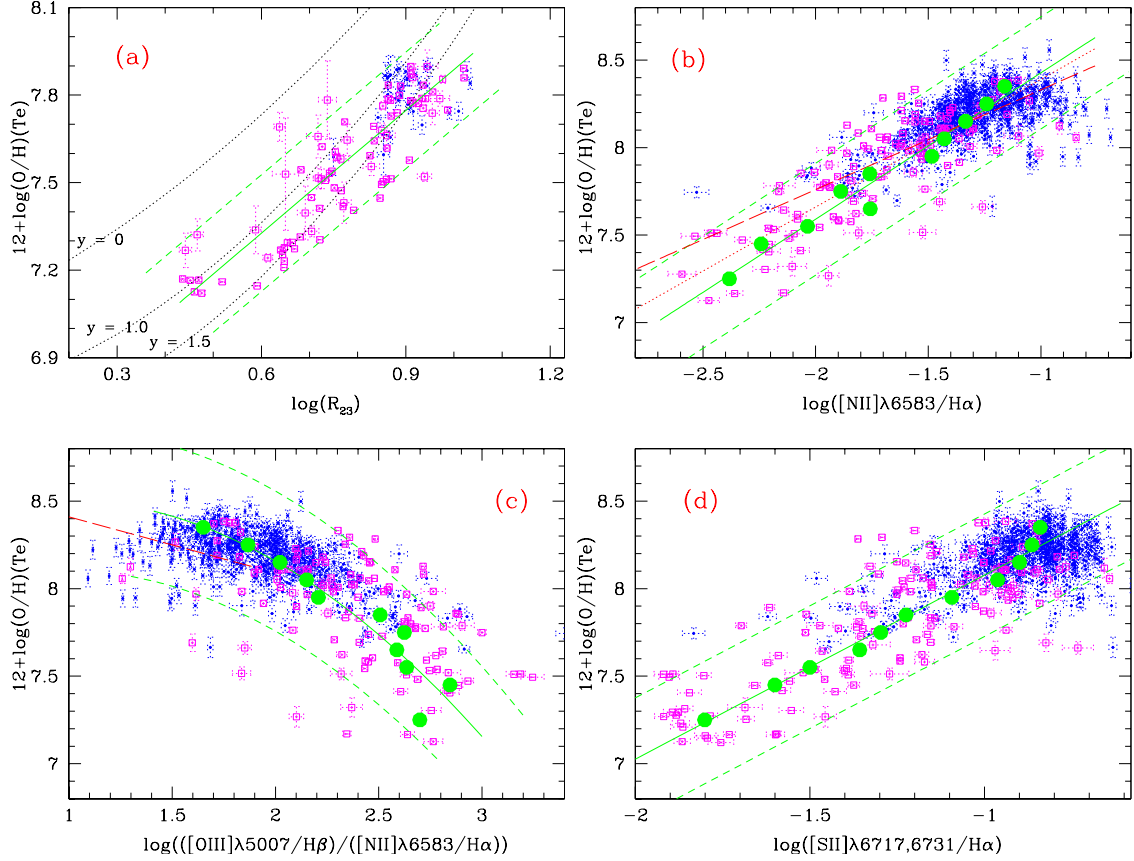}
\caption {$R_{23}$, N2, O3N2 and S2 indices for oxygen abundance calibrations
from the sample galaxies:
{\bf (a).} the $R_{23}$ index;
{\bf (b).} the N2 index;
{\bf (c).} the O3N2 index;
{\bf (d).} the S2 index.
The small filled points represent the SDSS galaxies and the open squares
refer to the samples from literature.
The points with uncertainties larger than 0.06\,dex in all these line ratios and
log(O/H) abundances have been excluded.
The solid line in Fig.(a) refers the linear least-squares fit
for all the 120 data points in the figure.
The three dashed lines refer the calibrations given by
Kobulnicky et al. (1999) for the three different cases of
$y$ (=log([O~{\sc iii}]/[O~{\sc ii}])).
In Fig.(b), the long-dashed and dotted lines refer to
the calibrations of PP04 and D02, respectively.
In Fig.(c), the long-dashed line refers to the calibration of PP04.
In Figs.(b-d),
the 11 large filled circles
represent the mean values in each bin of oxygen abundances (see text),
and the solid lines are the least-squares fits for these mean-value
points.
The two dashed lines show the 2$\sigma$ discrepancy from the fits
in all the four panels.
 }
\label{fig5}
\end{figure*}

We calculated the mean values of S2 and O/H abundances for the sample
galaxies in 11 bins, the bin widths are the same as we used for the
N2 and O3N2 calibrations. Again, the
objects with an higher uncertainty than 0.06\,dex on log(O/H) and
log([S~{\sc ii}]/H$\alpha$) are excluded in the calculations, which
leaves 607 data points. A least-squares linear fit to the mean-valued
points yields the relation:
\begin{equation}
\label{eqS2}
12 + {\rm log(O/H)} =9.128 + 1.051 \times S2,
\end{equation}
 with S2 = log([S~{\sc ii}]$\lambda\lambda$6717,6731/H$\alpha$),
which is shown by the solid line in Fig.~\ref{fig5}d.  The rms of the
data to the fit is about 0.176\,dex. The two dashed lines show the
discrepancy of 2$\sigma$.

\subsection{Add the P parameter}
\label{sec6.5}

The calibration of O/H vs. N2 is dependent on ionization parameter
(Fig.~\ref{fig4}b), as well the O3N2 and S2
indices (Figs.~\ref{fig4}c,d).
Therefore, to get more reliable O/H abundances,
 an excitation parameter could be included in the calibrations
which correlates strongly with ionization parameter and
 the hardness of the ionizing radiation field.
We use the $P$ parameter given by Pilyugin (2000, 2001a,b), where $P$
was defined as [O {\sc iii}]/([O {\sc ii}]+[O {\sc iii}]),
which is similar to [O {\sc iii}]/[O {\sc ii}], the excitation
parameter used by Kobulnicky et al. (1999) in calibrations.
We should notice that $P$ needs to be computed from the
dereddened emission lines.

\begin{figure}
\centering
\includegraphics[bb=154 316 441 700,width=8.0cm,clip]{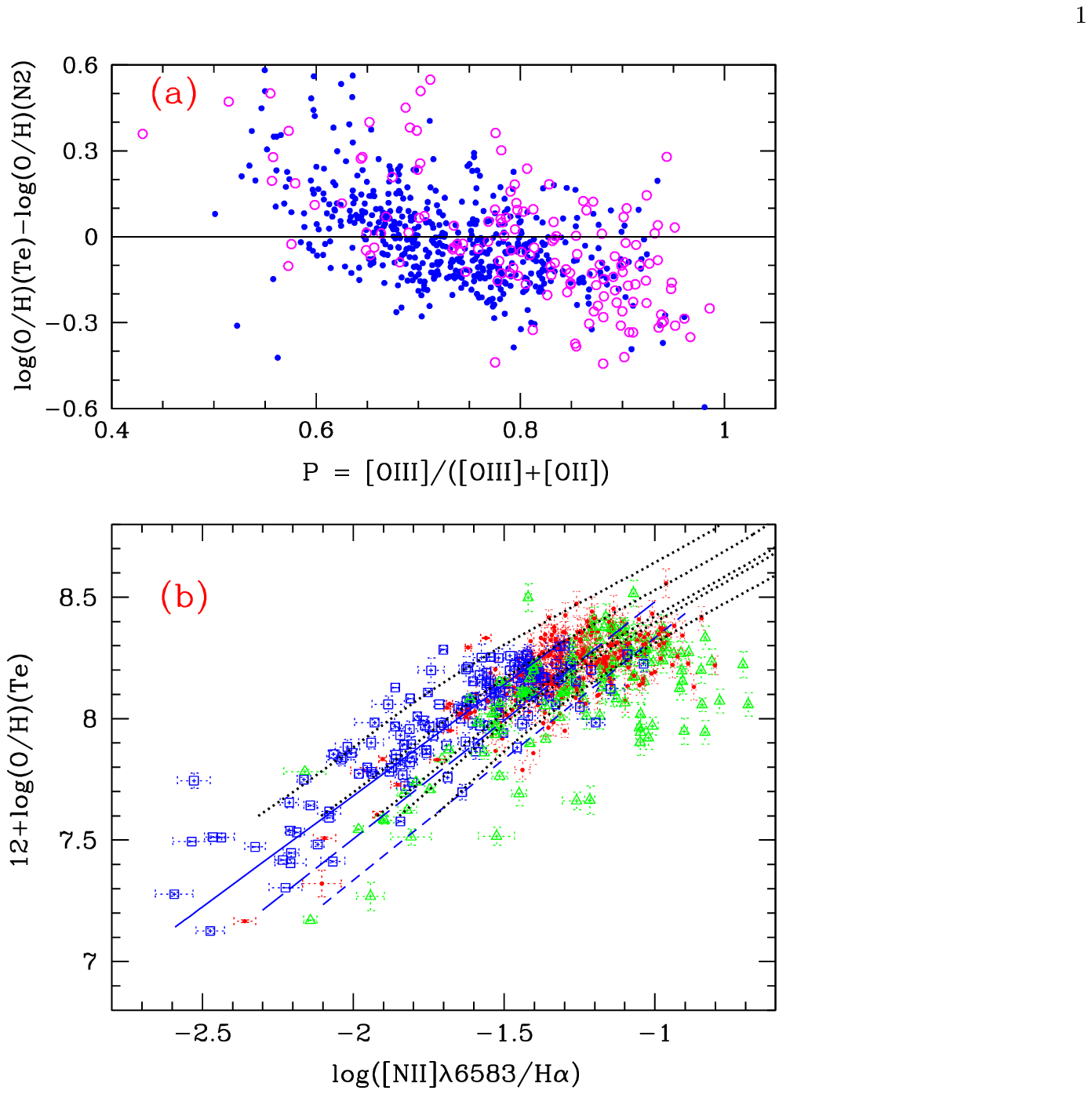}
\caption {{\bf (a).} The relation between log(O/H)$_{T_e}$-log(O/H)$_{N2}$
vs. $P$ for the sample galaxies, which shows that they can be
roughly divided into three subsamples with different $P$ parameters:
$P > 0.80$, $0.65 < P < 0.80$ and $P < 0.65$. The symbols are
the same as in Fig.~\ref{fig4}.
{\bf (b).}
The calibration of N2 index together with the $P$ parameter.
The blue open squares represent the samples with $P > 0.80$,
the red points represent those with $0.65 < P < 0.80$, and
the green triangles represent those with $P < 0.65$.
The solid, long-dashed and short-dashed lines refer to the
linear least-squares fits for each of the three subsamples,
respectively (Eq.~(\ref{eqN2P})).
The five dotted lines are the model results of KD02, from the right to the
left ones, the ionization parameter $q$ increases from the 5$\times
10^6$ to the 1$\times 10^7$, 2$\times 10^7$, 4$\times 10^7$ and
 8$\times 10^7$ cm s$^{-1}$ grids.
}
\label{fig6}
\end{figure}

In Fig.~\ref{fig6}a, we plot the (log(O/H)$_{T_e}$ - log(O/H)$_{N2}$) vs.
$P$ relations for the sample
galaxies, which shows that these galaxies can roughly be separated to be three
subsamples with $P<$0.65 (143 data points),
0.65$<P$$<$0.80 (292 data points) and $P>0.80$ (149 data points).
We derive the N2 calibrations for these three subsamples
individually.
 The analytical
calibrations of linear least-squares fits for the three subsamples with
different $P$ parameters are:
\begin{eqnarray}
\label{eqN2P}
12 + \rm log(O/H) & = & 9.514 + 0.916 \times N2, ~~~(P > 0.80), \nonumber \\
                  & = & 9.457 + 0.976 \times N2, ~~~(0.65 < P < 0.80), \nonumber \\
                  & = & 9.332 + 0.998 \times N2, ~~~(P < 0.65),
\end{eqnarray}
which are given as the solid, long-dashed and dashed lines in
Fig.~\ref{fig6}b, respectively.  The rms of the data to each fit are
0.150, 0.140, 0.189\,dex, respectively.  The dotted lines are the
photoionization model results of KD02.  It shows that the three
different $P$ values just correspond to the cases of
KD02 of $q=4\times 10^7$, $1\times
10^7$(as well $2\times 10^7$) and $5\times 10^6$ cm\,s$^{-1}$, respectively.

We have also tried to add $P$ as an additional parameter for the O3N2
and S2 indices.
The sample galaxies are separated into
 three sub-samples with
 $P$$<$0.60, 0.60$<$$P$$<$0.85 and $P$$>$0.85
to derive calibrations for these two indices. The basic results
follow the photoionization models of KD02 consistently.
 However, we find adding a $P$ parameter does not improve the fits much.
 Therefore, we won't present the two-parameter
calibrations for the O3N2 and S2 indices as figures here.
 We should notice that using $P$ with [N {\sc ii}]/H$\alpha$
 means that we need observations covering a much
 wider wavelength range, which might not be practical at high $z$.
It is not necessary to add the $P$ parameter to the $R_{23}$
calibration since the observational data in the $R_{23}$ vs. O/H relation
does not show obvious dependence on ionization parameter in contrast
to the lower-branch models (Fig.~\ref{fig4}a).

\subsection{The accuracies of the derived calibrations}
\label{sec6.6}

\begin{figure}
\centering
\includegraphics[bb=75 280 426 633,width=8.cm,clip]{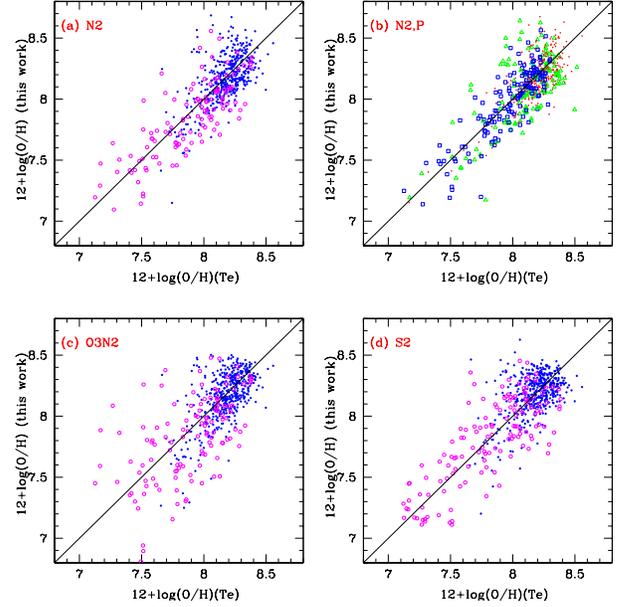}
\caption {Comparisons between the $T_e$-based O/H abundances with those
derived from strong-line ratios, including N2 (without and with $P$),
O3N2 and S2 indices:
({\bf a}). for the N2 index (without $P$, Eq.(\ref{eqN2}), Fig.~\ref{fig5}b);
({\bf b}). for the N2 index (with $P$, Eq.(\ref{eqN2P}), Fig.~\ref{fig6});
({\bf c}). for the O3N2 index (Eq.(\ref{eqO3N2}), Fig.~\ref{fig5}c);
({\bf d}). for the S2 index (Eq.(\ref{eqS2}), Fig.~\ref{fig5}d).
The rms values of the data to the equal-value lines
in each plot are 0.162, 0.158, 0.191 and 0.171\,dex,
respectively, they will increase to be
0.173, 0.167, 0.215 and 0.203\,dex if the data points with
uncertainties larger than 0.06\,dex in line indices and log(O/H) are included.
 The symbols in Figs.(a,c,d) are the same as in Fig.~\ref{fig4},
and the symbols in Fig.(b) are the same as in Fig.~\ref{fig6}b.}
\label{fig7}
\end{figure}

We have presented the least-squares fits for the relations of
12 + log(O/H)$_{T_e}$ and $N$2, $O$3$N$2 and $S$2 indices with and
without considering the $P$ excitation parameter.
To show the accuracies of these calibrations,
we plot in Figs.~\ref{fig7}a-d the comparisons between
log(O/H)$_{T_e}$ and those from line indices.  Figs.~\ref{fig7}a,c,d
show the comparisons with the calibrated abundances from N2, O3N2, S2
indices following Eqs.~(\ref{eqN2}-\ref{eqS2}),
which have been given as the solid lines
in Figs.~\ref{fig5}b-d, respectively. Fig.~\ref{fig7}b show the
calibrated abundances from the N2 index but adding the $P$ parameter
following Eq.~(\ref{eqN2P}).
 Adding the $P$ parameter has improved the N2 calibration.
Among these, the rms of the O3N2 and S2 calibrations are large, about
0.191 and 0.171\,dex respectively (Fig.~\ref{fig7}c,d).
The
calibration of (N2,$P$) case (Fig.~\ref{fig7}b) shows the lowest rms,
which is about 0.158, a bit lower than the rms in Fig.~\ref{fig7}a
and Fig.~\ref{fig3}e for the N2 calibration (0.162 and 0.160\,dex respectively).

\begin{figure}
\centering
\includegraphics[bb=18 158 591 491,width=8.cm,clip]{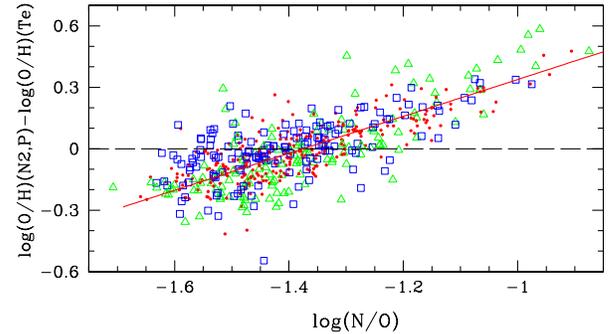}
\caption {Correlation between
the difference of log(O/H)$_{N2,P}$ and log(O/H)$_{T_e}$
and the log(N/O) abundance ratios of the sample galaxies.
The symbols are the same as in Figs.~\ref{fig6}b,\ref{fig7}b.
The solid line is the linear least squares fit for the relations
with a rms of 0.104\,dex.}
\label{fig8}
\end{figure}

 Some of the remaining scatter in log(O/H)$_{N2,P}$ may come from
the different star forming history and evolutionary status of the
sample galaxies.
Fig.~\ref{fig8} shows a clear
correlation between the log(O/H)$_{N2,P}$ - log(O/H)$_{T_e}$
and the log(N/O) abundance ratios of the sample galaxies,
which can be given as a linear least squares fit:
\begin{equation}
\label{N2pTe}
\rm log(O/H)_{N2,P} - log(O/H)_{T_e} = 0.902 \times log(N/O) + 1.240,
\end{equation}
with a rms of 0.104\,dex.
The corrected calibration of (N2,P) by considering this
correlation will provide much accurate oxygen abundance,
and the
rms of the data to the equal-value lines in Fig.~\ref{fig7}a and Fig.~\ref{fig7}b
will decrease to be
0.115\,dex, which is much less than the previous ones
(0.162 and 0.158\,dex, respectively).
This correlation can be used
to correct the derived oxygen abundances from strong-line ratio
when it is necessary.
 Nevertheless, to apply the N/O correction to
 12+log(O/H) abundances, we need to measure [O~{\sc ii}],
 [O~{\sc iii}] and H$\beta$ in addition
 to [N~{\sc ii}] and H$\alpha$, thus it will not always be practical.

 We should notice that both the $x$ and $y$ axis of
Fig.~\ref{fig8} depends on nitrogen abundance, which means that
the observed correlation could be spurious
 if the errors in log(N) are large.
However, the errors of the log(N/O) abundances of the sample
galaxies are quite small. It is about 0.011\,dex on average for
the MPA/JHU sample galaxies, which is expected since we only
select the objects with higher signal-to-noise ratio, i.e. larger
than 5$\sigma$, on the flux measurements of [N~{\sc ii}] (see
Sect.2.1). For other samples from literature, the errors of their
log(N/O) values are about 0.021\,dex on average. Therefore, this
correlation is robustly measured, and could be used
 to correct 12+log(O/H)$_{\rm N2,P}$ measurements if desired.  More
 importantly, the strong trend in Fig.~\ref{fig8} highlights the sensitivity of
 the N2 calibration to a galaxy's \emph{past} history of star
 formation, since it determines the present-day N/O ratio.

\section{Conclusions}

We compiled a large sample of 695 low-metallicity galaxies and H {\sc ii}
regions with [O~{\sc iii}]4363 detected, which consists of
531 galaxies from the SDSS-DR4 and 164 galaxies and H {\sc ii} regions
from literature.  We determined their
electron temperatures, electron densities, and then the $T_e$-based
oxygen abundances. The derived oxygen abundances of them are
7.1$<$12+log(O/H)$<$8.5.

The comparison between the $T_e$-based oxygen abundances and the
Bayesian metallicities obtained by the MPA/JHU group show that, for about
half of the MPA/JHU sample galaxies in this study, their O/H abundances
were overestimated by a factor of 0.34\,dex on average by using the
photoionization models (Charlot et al. 2006;
Tremonti et al. 2004; Brinchmann et al. 2004).
This is possibly related to how
  secondary nitrogen enrichment is treated in the models they used.

The $T_e$-derived oxygen abundances of these sample galaxies were
compared with those derived from other strong-line ratios, such as
$R_{23}$, $R_{23}-P$, $N$2 and $O$3$N$2 indicators.  The results show that
$R_{23}$ and $R_{23}-P$ methods will generally overestimate the oxygen
abundances of the sample galaxies by a factor of $\sim$0.20\,dex and
$\sim$0.06\,dex, respectively.
This 0.06\,dex discrepancy is quite small, and will {decrease to  be 0.025\,dex}
when the improved $R_{23}-P$ calibrations given by Pilyugin \& Thuan (2005)
are used. The
abundance calibration of N2 index, rather than O3N2 index, from PP04
can result in consistent O/H abundances with (O/H)$_{T_e}$ but with
a scatter of about 0.16\,dex.

From this large sample of 695 star-forming galaxies and H {\sc ii} regions,
we re-derive
abundance calibrations for $R_{23}$, N2, O3N2 and S2 indices on the basis of
their oxygen abundances derived from $T_e$. The N2, O3N2 and S2
indices monotonicly change following the increasing O/H abundances
from 12+log(O/H)=7.1 to 8.5,
which are consistent with the photoionization model results of KD02.
 $R_{23}$ is a good metallicity indicator
for the metal-poor galaxies with 12+log(O/H)$<$7.9.
 Nevertheless, $R_{23}$ does not follow the KD02 models well.

The scatter of the observational data may come from the
 differences in the ionizing radiation field.
 We add an excitation parameter $P$ to separate the
sample galaxies to be three sub-samples, and then re-derive the N2
calibrations with three different $P$ cases
($<$0.65, 0.65-0.80, $>$ 0.80), which
improves the calibration, confirmed by
 the decreasing rms of the data to the equal-value line,
 which decreases from 0.162
in  Fig.~\ref{fig7}a to 0.158 in Fig.~\ref{fig7}b.
But adding the $P$ parameter in the O3N2 and
S2 indices does not improve the calibrations.
However, some of the remaining scatter in the (N2,P) calibration may come from
the different star forming history and evolutionary status of the
galaxies, which can be confirmed by the correlation
between the log(O/H)$_{N2,P}$ - log(O/H)$_{T_e}$ and log(N/O)
of them (Fig.~\ref{fig8} and Eq.~\ref{N2pTe}).

In the future studies about calibrating oxygen abundances of galaxies,
when their $T_e$ values are not available, we can use the $R_{23}$, N2, O3N2 or
S2 indices, together with the $P$ parameter sometimes (if the wavelength
coverage is wide enough), to calibrate
their oxygen abundances. $R_{23}$ shows obvious correlation
with oxygen abundances for the low-metallicity galaxies with
12+log(O/H)$<$7.9.
 Comparing with O3N2 and S2, N2
is less affected by ionization parameter and more obviously
correlate with O/H abundances. Also it can be detected by the near
infrared instruments for the intermediate- and high-$ z$
star-forming galaxies.  The N2 index has helped to provide
important information on the metallicities of galaxies at
high-$z$, for example, Shapley et al. (2004) for a sample of 7
star-forming galaxies at 2.1$<$$z$$<$2.5, Shapley et al. (2005)
for a sample of 12 star-forming galaxies at $z\sim$1.0-1.5, and
Erb et al. (2006) from the six composited spectra of 87 rest-frame
UV-selected star-forming galaxies  at $z$$>$2 etc.  However,
considering the dependence of N2 (as well the (N2,P)) calibration
on log(N/O) shown as Fig.~\ref{fig8}, we should keep in mind that
the resulted log(O/H) abundances from N2 may involve the specific
history of N-enrichment in the galaxies. We may worry a bit about
comparing the low and high-$z$ observations using [N~{\sc ii}] for
this reason.

Given the good consistency of our N2 and S2 abundance
calibrations with the models of KD02 at low metallicity (see
Fig.~\ref{fig4}b,d), it may be appropriate to linearly
extrapolate our calibrations up to the high metallicity region,
up to 12+log(O/H)$\sim$9.0 and 8.8 respectively. However, this
extrapolation on N2 will result in lower log(O/H) abundance
than the one directly derived from the calibration of the
metal-rich SDSS galaxies obtained by Liang et al. (2006), which
may be due to the discrepancy between the $T_e$- and the
$R_{23}$-based metallicities. However, the $T_e$-based O3N2
calibration cannot be directly extrapolated to the metal-rich
region since the slopes of the correlation relations between
log(O/H) vs. O3N2 are not the same in the high-metallicity
branch as in the low-metallicity branch (Fig.~\ref{fig4}c).
$R_{23}$ parameter cannot provide reliable oxygen abundances
for the galaxies in the metallicity turn-over region with
7.9$<$12+log(O/H)$<$8.5.

\section*{Acknowledgments}
  We thank the referee for many valuable and wise suggestions and
  comments, which help us in improving well this work.
  We thank Stephane Charlot for the interesting discussions about
  their models to estimate the oxygen abundances of the
  SDSS galaxies.
  S.Y.Y. and Y.C.L. thank Jing Wang and Caina Hao for the interesting
  discussions about the SDSS database.
  S.Y.Y thanks Yongheng Zhao and the LAMOST group kindly provide
  their office for staying.
  This work was supported by the Natural Science Foundation of China
 (NSFC) Foundation under No.10403006.

\end{document}